\documentclass[final]{elsart3p}

\usepackage{graphicx}

 \journal{J. Magn. Magn. Mater.}
\usepackage{dcolumn}
\usepackage{bm}

\begin{document}

\begin{frontmatter}

\title{Role of defects on the electronic and magnetic properties of CrAs/InAs and CrAs/CdSe half-metallic interfaces}

\author{I. Galanakis\corauthref{cor}}\ead{galanakis@upatras.gr}
\author{ and I. Lekkas}

\address{Department of Materials Science, School of Natural
  Sciences, University of Patras,  GR-26504 Patra, Greece}
 \corauth[cor]{Corresponding author. Phone +30-2610-969925,
Fax +30-2610-969368}

\begin{abstract}
We present an extended study of single impurity atoms at the
interface between the half-metallic ferromagnetic zinc-blende CrAs
compound and the zinc-blende binary InAs and CdSe semiconductors
in the form of very thin multilayers. Contrary to the case of
impurities in the perfect bulk CrAs studied in [I. Galanakis and
S.G. Pouliasis, J. Magn. Magn. Mat. 321 (2009) 1084] defects at
the interfaces do not alter in general the half-metallic character
of the perfect systems. The only exception are Void
 impurities at Cr or In(Cd) sites which lead, due to the lower-dimensionality
 of the interfaces with respect to the bulk CrAs, to a shift of the
  $p$  bands of the nearest neighboring As(Se)
 atom to higher energies and thus to the loss of the half-metallicity. But Void impurities are
 Schottky-type and should exhibit high formation
 energies and thus we expect the interfaces in the case of thin
 multilayers to exhibit a robust half-metallic character.
\end{abstract}

\begin{keyword}
Electronic structure \sep Half-metals \sep CrAs

\PACS 75.47.Np \sep 75.50.Cc \sep 75.30.Et
\end{keyword}
\end{frontmatter}

\section{Introduction}\label{sec1}

The discovery of Giant Magnetoteresistance in 1998 by the groups
of Fert and Gr\"unberg led to new reading heads for hard disks
\cite{Fert,Grunberg}. Moreover for the first time, a device based
on magnetic phenomena replaced a conventional electronics device
based on the movement of the electrons charge and thus opened the
way to the field of spintronics or magnetoelectronics. The aim is
to replace conventional electronics with new devices where
magnetism plays a central role leading to smaller energy
consumption. Several architectures have been proposed
\cite{Zutic,Zabel} but only in 2009 Dash and collaborators managed
to inject spin-polarized current from a metallic electrode into
Si, which is a key issue in current research in this field.
showing that spintronic devices can be incorporated into
conventional electronics \cite{Dash}.

In order to maximize the efficiency of spintronic devices, the
injected current should have as high spin-polarization as possible
\cite{Zutic,Zabel}. To this respect half-metallic compounds have
attracted a lot of interest (for a review see reference
\cite{Katsnelson}). These alloys are ferromagnets where the
majority spin channel is metallic while the minority-spin band
structure is that of a semiconductor leading to 100\%\
spin-polarization of the electrons at the Fermi level and thus to
possibly 100\%\ spin-polarized current into a semiconductor when
half metals are employed as the metallic electrode. The term
half-metal was initially used by de Groot et al in the case of the
NiMnSb Heusler alloy \cite{Groot}.

Ab-initio (also known as first-principles) calculations have been
widely used to explain the properties of these alloys and to
predict new half-metallic compounds. An interesting case is the
transition-metal pnictides like CrAs and MnAs. Akinaga and
collaborators found in 2000 that when a CrAs thin film is grown on
top of a zinc-blende semiconductor like GaAs, the metallic film
adopts the lattice of the substrate and it crystallizes in a
meta-stable half-metallic zinc-blende phase \cite{Akinaga}
structure. Later CrAs was successfully synthesized in the
zinc-blence structure in the form of multilayers with GaAs
\cite{Mizuguchi} and other successful experiments include the
growth of zinc-blende MnAs in the form of dots \cite{Ono} and CrSb
in the form of films \cite{Zhao,Li}.

Experiments agree with predictions of ab-initio calculations
performed by several groups
\cite{MavropoulosZB,GalaZB,calculations,Shirai}. In the case of
the half-metallic ferromagnets like CrAs or CrSe, the gap in the
minority-spin band arises from the hybridization between the
$p$-states of the $sp$ atom and the triple-degenerated $t_{2g}$
states  of the transition-metal and as a result the total
spin-moment, $M_t$, follows the Slater-Pauling (SP) behavior being
equal in $\mu_B$ to $Z_t-8$ where $Z_t$ the total number of
valence electrons in the unit cell \cite{MavropoulosZB}. Recently
theoretical works have appeared attacking also some crucial
aspects of these alloys like  the exchange bias in
ferro-/antiferromagnetic interfaces \cite{Nakamura2006}, the
stability of the zinc-blende structure \cite{Xie2003}, the
dynamical correlations \cite{Chioncel2006}, the interfaces with
semiconductors \cite{InterGala,Interfaces}, the exchange
interaction \cite{Sasioglu-Gala}, the emergence of half-metallic
ferrimagnetism \cite{Rapid} and the temperature effects
\cite{MavropoulosTemp}. An extended overview on the properties of
these alloys can be found in reference \cite{Review}.

\begin{figure}
\begin{center}
\includegraphics[width=\columnwidth]{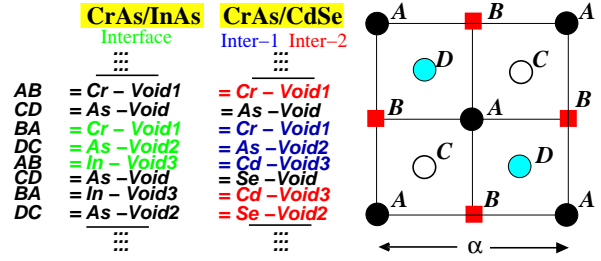}
\end{center} \caption{(Color online) Schematic representation of the structure of the
multilayers under study. We have assumed that the growth direction
of  the zinc-blende structure is the (001) and the period consists
of four CrAs and four InAs(CdSe) layers. Each plane contains atoms
of a single chemical type plus a void site. We denote as Void1 the
vacant site in the same layer with the Cr atoms at the interfaces,
Void2 in the interfacial As(Se) layer and Void3 in the interfacial
In(Cd) layer. The distance between two consecutive plane is
$\frac{1}{4}$ of the lattice constant. Note that in the case of
the CrAs/CdSe we have two non-equivalent interfaces: (i) when the
sequence of the atoms is ...-Cr-As-Cd-... denoted as CrAs/CdSe-1
and (ii) when the sequence is ...-Cr-Se-Cd-... denoted as
CrAs/CdSe-2. Finally we should note that we have assumed the
lattice constant of the two semiconductors (0.606 nm).
\label{fig1}}
\end{figure}

\begin{table*}
\centering \caption{Atom-resolved spin magnetic moments in $\mu_B$
for the perfect CrAs and CrSe alloys and for the three perfect
multilayers under study. For the position of each atom in the
multilayer please refer to figure \ref{fig1}.}
 \begin{tabular}{l|ccccc} \hline \hline
Perfect Systems&  CrAs(Bulk) &  CrSe(Bulk)&  CrAs/InAs &
CrAs/CdSe-1 & CrAs/CdSe-2 \\
Cr & 3.267 &   3.825 &   3.248&    3.117 & 3.462 \\
Void1 & -0.029& 0.049&    -0.008&   -0.025& -0.002 \\
As(Se) [CrAs layer]  & -0.382&   -0.103(Se)&  -0.383& -0.388 & \\
As(Se) [Interface] &&&  -0.209&   -0.365 &  -0.093(Se)\\
 As(Se) [SemiConducting layer]&   &&  -0.066& &      -0.048(Se) \\
Void2 & 0.080&    0.162 &   0.043&    0.027&    0.056\\
 In(Cd)&& &-0.008&   -0.045&   0.003 \\
 Void3&&& -0.022 &-0.033 & -0.001
 \\\hline \hline
\end{tabular}
\label{table1}
\end{table*}

Recently we have published a work on the role of defects in the
case of half-metallic CrAs, CrSb and CrSe alloys crystallizing in
the zinc-blende structure and adopting the lattice constant of
InAs binary semiconductor of 0.606 nm \cite{Pouliasis}. All
defects under study in this reference, with the exception of void
impurities at Cr and $sp$ sites and Cr impurities at $sp$ sites,
were found to induce new states within the gap and the Fermi level
can be pinned within these new minority states destroying the
half-metallic character of the perfect  bulk compounds. These
impurity states are localized in space around the impurity atoms
and very fast the bulk behavior is regained. But in realistic
devices interfaces with semiconductors will occur and the
interaction of these bulk-impurity states with interface states
can destroy the spin-polarization of the injected current.
Ab-initio calculations show that interfaces are in principle
half-metallic for pnictides containing Cr or V atoms even when the
semiconductor is a II-VI like CdSe and not a III-V one like InAs,
and no interface states occur \cite{InterGala}. But impurities at
the interfaces can induce new impurity-interfaces states within
the minority spin-gap which can couple to the bulk impurity states
and lead to  loss of the half-metallic character.

In this communication we expand our previous study to cover also
the case of impurities at interfaces. We have decided to consider
CrAs as the half-metallic spacer since it is the most-widely
studied transition-metal zinc-blende pnictide and both InAs and
CdSe as the zinc-blende semiconducting spacer to cover both the
case of III-V and II-VI semiconductors. In and Cd atoms, as well
as As and Se atoms, are in the same row of the periodic table and
they have one valence electron difference and thus both alloys
have the same lattice constant of 0.606 nm and for this value of
the lattice constant bulk CrAs shows a Fermi level exactly in the
middle of the minority-spin gap making clear the appearance of
impurity states. For our calculation we employ the
 the Korringa-Kohn-Rostoker method (KKR) method \cite{Pap02} as in the case of
 the perfect bulk \cite{MavropoulosZB} and interfaces
 \cite{InterGala} and we have treated the impurities as in
 references \cite{Pouliasis,imp}.

To model the interface we have considered a multilayer with a
period of 4 CrAs and 4 InAs(CdSe) monolayers (MLs) which is in
accordance with the experimental data in reference
\cite{Mizuguchi} and a schematic illustration is shown in figure
\ref{fig1}.  As growth direction we have chosen the (001) and thus
its layer is made up of atoms of a single chemical type. Moreover
at each layer we have considered a void to describe better the
space in our calculation. In the case of CrAs/InAs multilayers all
interfaces are equivalent while in CrAs/CdSe case we have two
non-equivalent interfaces; the interface where As is between the
Cr and Cd layers which we denote as Interface-1 and the interface
where Se separates the Cr and Cd layers (Interface-2). Moreover we
should note that the distance between two successive layers is 1/4
of the lattice constant and thus atoms in successive  layers are
nearest atoms while within the same atomic layer the nearest atoms
are second neighbors. We have also denoted as Void1 the vacant
site at the Cr interface layer, Void2 in the As(Se) interface
layer and Void3 in the In(Cd) interface layer. We have considered
all possible impurities and in section \ref{sec2} we present the
case of perfect bulk and mulitilayers while in section \ref{sec3}
we present the properties of the Cr impurities at various sites at
the interface, in section \ref{sec4} the case of As(Se)
impurities, in section \ref{sec5} the case of voids and in section
\ref{sec6} the case of In(Cd) impurities. Finally in section
\ref{sec7} we summarize and conclude. Throughout the discussion we
compare our results with the case of impurities in perfect bulk
CrAs presented in reference \cite{Pouliasis}. In the case of the
CrAs/CdSe-2 interface the Cr atoms have As atoms from one side and
Se atoms from the other side but as it was shown in
\cite{Pouliasis} both CrAs and CrSe show similar behavior with
respect to the defects-induced states.

\begin{figure}
\includegraphics[width=\columnwidth]{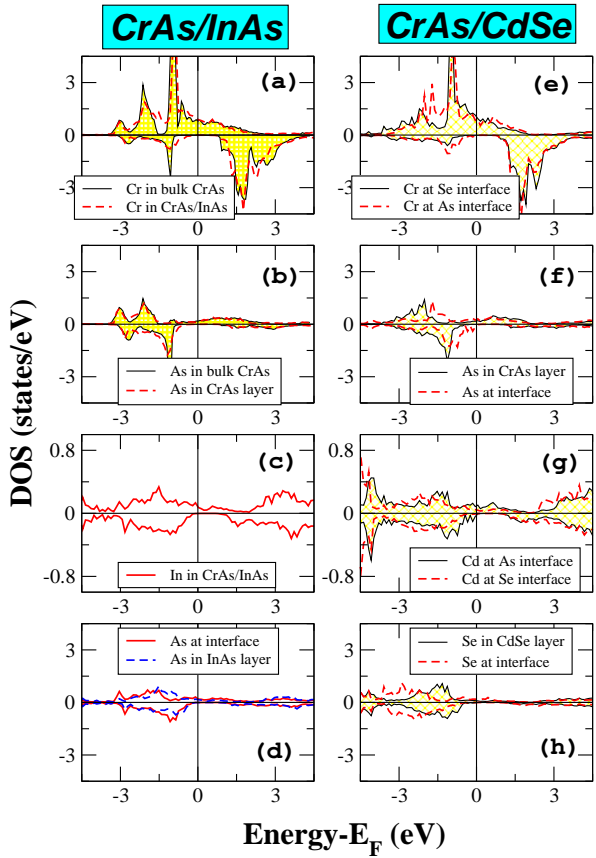}
\caption{(Color online) Atom-resolved density of states (DOS) for
the bulk CrAs and the perfect multilayers under study presented in
figure \ref{fig1}. Panels (a), (b), (c) and  (d) refer to the
CrAs/InAs multilayer and (e),(f), (g) and (h) to the CrAs/CdSe
multilayer. In panel (a) and (b) we have also included the DOS of
Cr and As atoms in perfect bulk CrAs for comparison. Note that in
the case of the CrAs/CdSe we have two interfaces one containing As
one containing Se atoms (see figure \ref{fig1} for explanation).
The Fermi energy has been set as the zero of the energy axis.
Positive DOS values correspond to
 the spin-up (majority-spin) electrons and negative values to the spin-down
 (minority-spin) electrons.\label{fig2}}
\end{figure}

\section{Perfect systems}\label{sec2}

We will start our discussion from the perfect systems. In table
\ref{table1} we have gathered the atom-resolved spin magnetic
moment in $\mu_B$ for the three interfaces and for the bulk CrAs
and CrSe alloys. Notice that for the bulk systems, where we have
two vacant sites, we denote as Void1  the vacant site with the
same symmetry as the Void1 site at the interface and the same
stands for the Void2 site. Bulk CrAs has 11 valence electrons per
unit cell while CrSe has 12 valence electrons and thus
half-metallic CrAs has a total spin moment of 3 $\mu_B$ and CrSe
of 4 $\mu_B$. As it was shown in \cite{MavropoulosZB} Cr atoms in
CrSe accommodate half of the extra electron in the majority-spin
band leading to a larger Cr-spin moment which now reaches about
3.8 $\mu_B$ instead of $\sim$3.3 $\mu_B$ in CrAs. The other half
electron is accommodated in the majority-spin band of the Se and
Void2 atoms as it easily deduced from the spin moments presented
in table \ref{table1} (Void2 is not a real atom but orbitals from
the nearest Cr atoms penetrate in the Void2 site).

In the case of the CrAs/InAs multilayer, both Cr atoms have
exactly the same nearest environment as in bulk CrAs and almost an
identical spin magnetic moment. The same is true also for the As
atom within the CrAs layer (As layer between the two Cr layer as
shown in figure \ref{fig1}) which has four Cr and four Void1 sites
as nearest neighbors as in the bulk CrAs and exactly the same spin
magnetic moment of -0.38 $\mu_B$. As at the interface layer has
one Cr layer from one side an In layer from the other side. In
atoms have two valence electrons which mainly are of $s$ character
and have no contribution to the In spin magnetic moment. The spin
moment at the In sites is induced by the As atoms since
$p$-orbitals of As penetrate in the In sites. Thus As atoms at the
interface show a smaller absolute value of the spin magnetic
moment since now they have  only two instead of four Cr nearest
neighbors (the spin-moment at As is induced by the Cr atoms
through the hybridization of the As-$p$ and Cr-$t_{2g}$ states)
and In atoms at the interface show a negligible spin moment
induced by the As interface atoms which in reality is the
reflection of the Cr spin moment. As atoms within the
semiconducting layer have four In atoms as nearest neighbors and a
very small spin magnetic moment.

In the case of the CrAs/CdSe multilayer we have two inequivalent
interfaces and the two Cr layer are no-more equivalent. The Cr
atoms at the As interface behave similarly to bulk CrAs although
the spin moment is somewhat smaller while in the case of the Se
interface Cr atoms have two As and two Se atoms as nearest
neighbors exhibiting a spin magnetic moment of about 3.5$\mu_B$ a
mean value between the bulk CrAs and CrSe atoms. Within the CrAs
layer As shows a spin magnetic moment similar to the CrAs/InAs
multilayer. At the 1$^\mathrm{st}$ interface As atoms show similar
behavior to the CrAs/InAs case although their spin moment is
larger in absolute value since Cd atoms have one valence electron
less that In ones and the leakage of charge from the As interface
atoms to the Cd ones is smaller than to the In ones. In the
2$^\mathrm{nd}$ interface Se atoms have one electron more than As
one and thus the Se-$p$ orbitals are less polarized from the
Cr-$t_{2g}$ ones and show a smaller spin magnetic moment by a
factor of four. Within the CdSe semiconducting layer the Se atoms
behave like the As atoms in the InAs layer in the first interface
and show a small value of the spin moment. Finally Cd atoms show a
negligible spin-magnetic moment as the In ones.

In figure \ref{fig2} we have gathered the density of states (DOS)
in states/eV for both multilayers under study and the bulk CrAs.
In the left column we present the case of the CrAs/InAs
multilayer. Cr atoms at the interface show a similar DOS to the
bulk case with a very small weight of occupied minority-spin
states while in the majority band both $e_g$ and a fraction of the
three $t_{2g}$ states are occupied. The gap is smaller in the case
of the multilayer since the occupied minority-spin states are
broader in energy and this is reflected to all other atoms. As
atoms within the CrAs layer show an almost identical DOS to the
bulk CrAs case as was also the case for the spin magnetic moments.
As we mote to the interface and then to the InAs layer the shape
of the As states changes due to the different local environment
and the bands move towards the Fermi level. The states shown are
the $p$ states since the $s$-states of all atoms are located at
about -9 to -12 eV below the Fermi level and are not shown in the
figures. Note that for the In atoms we have used a different scale
in the vertical axis. In the case of the CrAs/CdSe multilayer the
half-metallicity is again present as for the CrAs/InAs layer but
due to the character of the II-VI semiconductor the gap in the
minority-spin band is smaller and the Fermi level is near the
left-edge of the gap as it is clearly seen in the case of the Cr
DOS. The reduction of the gap-width is larger for the As than the
Se interface and this agrees with the fact that the bulk CrSe
presents a larger gap than the bulk CrAs due to the larger
electronegativity of the Se atom as it was shown in reference
\cite{MavropoulosZB}. As a consequence the states of the As atoms
at the first interface have moved closer to the Fermi level in a
rigid way with respect to the As atoms in the CrAs layer. Cd atoms
at both interfaces show similar DOS to the In atoms in the
CrAs/InAs interface. Se atoms within the CdSe layer show a similar
shape to the As atoms in the InAs layer while Se atoms at the
second interface show states deeper in energy than the As atom in
the first interface in agreement with the larger gap-width shown
by the Cr atoms in the second interface with respect to the first
CrAs/CdSe interface.
\begin{table*}
\centering \caption{Atom-resolved spin magnetic moments for the
case of Cr impurity atoms at interfacial As(Se) and Void2 sites in
the case of bulk CrAs and the three interfaces under study. "imp"
stands for impurity and "1st" stands for nearest neighbor atoms
(similar for 2nd and 3rd). We present results for both cases of
coupling of the Cr impurity spin moment with respect to the spin
moments of the  Cr nearest-neighbors (ferromagnetic-FM and
antiferromagnetic-AFM cases). Note that As(Se) atoms can be found
in the CrAs layer [CrAs], the interfaces [Inter] and the
semiconductiong layer [SC]. }
 \begin{tabular}{l|c|c|c|c|c|c|c|c|c} \hline \hline
\underline{\bf Cr at As(Se) site} & \multicolumn{2}{c|}{BULK CrAs}
& \multicolumn{2}{c|}{CrAs/InAs} &
\multicolumn{2}{c|}{CrAs/CdSe-1}&
\multicolumn{2}{c|}{CrAs/CdSe-2}\\
& \underline{FM} &  \underline{AFM} & \underline{FM} &
\underline{AFM} & \underline{FM} &
\underline{AFM} & \underline{FM} &  \underline{AFM}  \\
 Cr-imp & 4.388 & -3.690 & 4.334 & -3.802 & 4.248& -4.058 & 4.335 & -3.988 \\
 Cr-1st & 3.602 & 3.364 & 3.704 & 3.425 & 3.543 & 3.343 & 3.735 & 3.485 \\
 In(Cd)-1st & & & 0.043 & -0.067 & 0.044 & -0.108 & 0.063 & -0.104 \\
 As [CrAs]-3rd & -0.405 & -0.405 &   -0.412 & -0.413 & -0.418 & -0.418 &
 -0.025 & -0.074\\
As(Se) [Inter]-3rd & & & -0202 &-0.237 & -0.305 & -0.352 &-0.105(Se) & -0.110(Se) \\
As(Se) [SC]-3rd & & & -0.025 & -0.067 & -0.032(Se) & -0.059(Se) &-0.417(Se) & -0.418(Se) \\
\hline \hline \underline{\bf Cr at Void2 site} &
\multicolumn{2}{c|}{BULK CrAs} & \multicolumn{2}{c|}{CrAs/InAs} &
\multicolumn{2}{c|}{CrAs/CdSe-1}&
\multicolumn{2}{c|}{CrAs/CdSe-2}\\
& \underline{FM} &  \underline{AFM}  &\underline{FM} &
\underline{AFM} & \underline{FM} &
 \underline{AFM} & \underline{FM} &  \underline{AFM}  \\
Cr-imp & 3.901 & -2.908 &  3.821 & -3.262& 3.729& -3.263&   4.171 & -3.551 \\
 Cr-1st & 3.419 & 3.267& 3.412 &3.225 & 3.300 &3.103 & 3.467 &3.335 \\
 In(Cd)-1st & & & 0.014 & -0.176 & -0.118 &-0.030 &  0.023 & -0.038\\
 As [CrAs]-2nd &  -0.341 &-0.285 & -0.364 &-0.323 &  -0.371 & -0.330 &   -0.470 &-0.371 \\
 As(Se) [Inter]-2nd & & &  -0.167 &-0.141 & -0.308 &-0.214 & -0.102(Se) & -0.104(Se) \\
 As(Se) [SC]-2nd &  & &    -0.009 &-0.039&  -0.013(Se) &-0.033(Se) & -0.033(Se) & -0.026(Se)\\
\hline \hline
\end{tabular}
\label{table2}
\end{table*}

\begin{figure}
\includegraphics[width=\columnwidth]{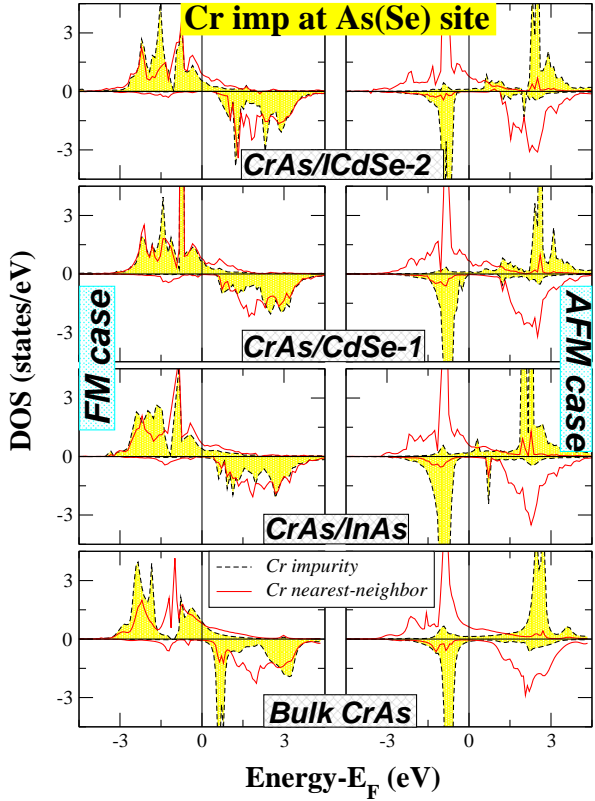}
\caption{(Color online)   Cr-resolved DOS for the case of Cr
impurity atoms at interfacial As(Se) sites at in the case of the
ferromagnetic (FM) [left panel] and antiferromagnetic coupling
(AFM) [right panel) between the Cr impurity atom and its Cr
nearest-neighbors (see text for details). Details as in figure
\ref{fig2}. \label{fig3}}
\end{figure}

\section{Cr impurities}\label{sec3}

We will start our discussion from the case of Cr impurities. The
first case under study is when Cr impurity atoms are located at
As(Se) or Void2 sites at the interface. With respect to the
perfect Cr sites where the first neighbors were four sp atoms and
four voids, the Cr impurity atoms have now as nearest neighbors
two Cr atoms and two In(Cd) atoms together with two Void1 and two
Void3 sites. Thus the local environment of the Cr impurities is
fundamentally different. In the case of Cr impurities at As and
Void2 sites in the bulk, we have shown that the spin magnetic
moment could be either ferromagnetically (FM) coupled to the spin
moment of the Cr atoms at the perfect sites or
antiferromagnetically (AFM) coupled with the latter one being the
energetically favorable case \cite{Pouliasis}. The AFM was
occurring due to the short distance between the Cr impurity atoms
and its four nearest Cr neighbors (Cr as well as Mn atoms are know
to have antiparallel spin moments under a critical distance). In
the case of the interfaces under study the Cr impurity atoms have
now two instead of four nearest Cr neighbors as in the bulk. But
it seems that this is enough to have again two stable solutions, a
FM and an AFM one, depending on the starting potential, for both
Cr at interfacial As(Se) and Void2 sites. The AFM solution seems
energetically more favorable although we have not converged the
energetics of the defects (for the reason see discussion in
reference \cite{Pouliasis}).

In figure \ref{fig3} we have gathered the atom-resolved DOS for
the Cr impurity atoms and its nearest Cr neighbors for all three
interfaces under study and bulk CrAs for the case of Cr impurity
at As(Se) site and for both the FM and AFM cases. Results are
similar when the Cr impurity is located at the Void2 site and thus
are not shown here. Also results are in all cases similar to the
bulk one with minor changes at the position of the picks due to
the variation in the environment surrounding the impurities. In
the FM case the impurities show almost zero occupied minority-spin
states while the Cr nearest neighbors show a very small weight of
minority-spin states below the Fermi level. In all cases the
half-metallic character of the interface is not affected. In the
AFM case the Cr impurities exhibit a large spin-splitting of the
bands and almost all the spin-down states are occupied while
almost all spin-up states are empty. This leads to a large gap in
the spin-down band between the spin-down occupied bands of the
impurity atom and the unoccupied spin-down bands of the Cr nearest
neighbors and the half-metallic character of the interface is not
affected. In table \ref{table2} we have gathered the atomic spin
magnetic moments for all cases mentioned above. We can see that
spin moments behave similar to the case of the Cr impurities at As
or Void2 sites in the perfect bulk. In the FM case the Cr impurity
atoms have a larger spin moment with respect to the Cr atoms in
the perfect systems since the weight of the minority occupied
states has vanished as a result of the hybridization with its
first neighbors. The Cr atoms which are first neighbors of the
impurity atoms present spin moments almost identical to the
perfect compounds in table \ref{table1}. In the AFM case although
it seems from the DOS that all five spin-down $d$-states of the
impurity atom are occupied the spin moment is less than 5 $\mu_B$
since some spin-up states also appear below the Fermi level
reflecting the band-structure of the Cr nearest neighbors. We
should note that spin moments are slightly smaller when the
impurity is located at a Void2 instead of a As(Se) site because
although the nearest-neighbors are the same, the next-nearest and
further neighbors are different. We should finally note here that
the AFM case is the most interesting case for applications since
the occurrence of half-metallic ferrimagnetism leads to smaller
external fields and exhibiting smaller energy losses with respect
to ferromagnets.

\begin{table}
\centering \caption{Atom-resolved spin magnetic moments in $\mu_B$
for the case of Cr impurity atoms at interfacial Void1, In(Cd) and
Void3 sites in the case of the three interfaces under study.
Notation as in table \ref{table2}. }
 \begin{tabular}{l|c|c|c} \hline \hline
\underline{\bf Cr at Void1 site}  &  CrAs/InAs & CrAs/CdSe-1 &
CrAs/CdSe-2 \\
Cr-imp & 3.522&  3.363 & 3.686 \\
 As(Se) [CrAs]-1st & -0.300 & -0.299 & -0.296 \\
 As(Se) [Inter]-1st & -0.151& -0.275& -0.042 \\
   Cr [Inter]-2nd & 3.448&   3.352 & 3.622\\ \hline
\underline{\bf Cr at In(Cd) site}  &   CrAs/InAs & CrAs/CdSe-1 &
CrAs/CdSe-2 \\ Cr-imp &  3.185 & 3.311&     4.015 \\
 As(Se) [SC]-1st & -0.157 & -0.095& -0.147 \\
  As(Se) ([Inter]-1st &-0.324 & -0.406& -0.254 \\
  Cr [Inter]-3rd &  3.328& 3.252 & 4.219 \\ \hline
\underline{\bf Cr at Void3 site}  &   CrAs/InAs & CrAs/CdSe-1 &
CrAs/CdSe-2 \\
Cr-imp & 3.508& 3.501&    3.833\\
 As(Se) [SC]-1st &-0.024 &-0.004 & 0.009 \\
  As(Se) [Inter]-1st & -0.135 & -0.275 & -0.025 \\
   Cr [Inter]-2nd &   3.478 & 3.384 & 3.635 \\
 \hline \hline
\end{tabular}
\label{table3}
\end{table}

\begin{figure}
\includegraphics[width=\columnwidth]{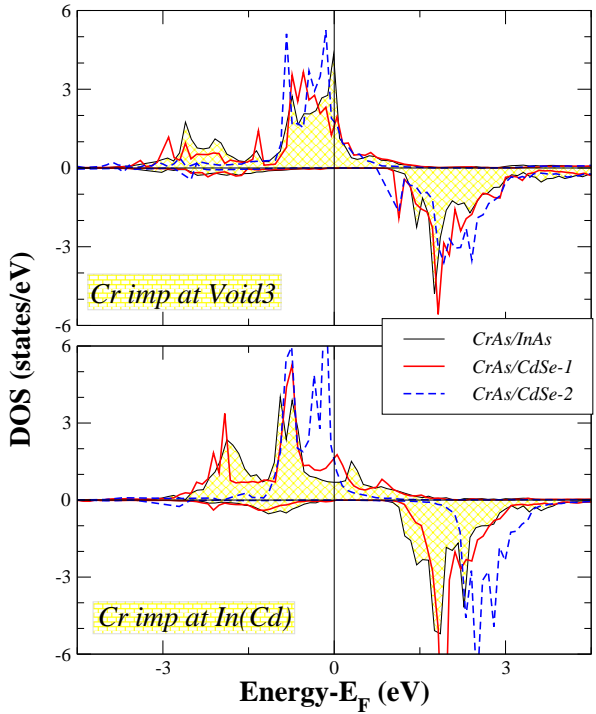}
\caption{(Color online) Density of states of the Cr impurity atom
at Void3(upper panel) and In(Cd) (lower panel) interfacial sites
for all the interfaces under study. Details as in figure
\ref{fig2}. \label{fig4}}
\end{figure}

The next case under study is when the Cr impurity is located at a
Void1 site. Although the Void1 site is located in the same layer
as the Cr atoms, they are second neighbors having the same nearest
neighbors (four sp atoms and four voids). Thus we do not expect Cr
impurities at Void1 sites to alter the electronic and magnetic
properties of the interface similarly to the situation of Cr
impurities at Void1 sites in the bulk CrAs \cite{Pouliasis}. The
DOS is similar to the Cr atoms in the perfect systems and thus we
do not present them since the gap is almost unaltered and the
half-metallicity is not affected. The spin magnetic moment of the
Cr impurity atoms which are shown in table \ref{table3} are
slightly larger than the Cr atoms in the perfect system by about
0.2 $\mu_B$. The Cr nearest-neighbors have spin moment almost
identical to the Cr impurities and thus part of the charge of the
Cr impurity atoms is used to increase the spin moment of the Cr
second neighbors with respect to the perfect interfaces.

The final case under study in this section is when the Cr impurity
occurs at the In(Cd) sites at the interface or the Void3 sites
which are located in the same layer with the In(Cd) ones. Although
one could think that these impurities should destroy the
half-metallic character of the interface, this is not true. Cr
impurities at these sites have as nearest neighbors two As(Se)
atoms in the semiconducting space and two As(Se) site at the
interface. Thus their local environment is quite close to the one
of the Cr atoms at the perfect interface sites. As a result the
half-metallic character is not destroyed as shown by the DOS in
figure \ref{fig4}. When the impurity is located at a Void3 site,
there is almost no-occupied spin-down states while when the
impurity is located at the In(Cd) site the weight of the occupied
spin-down states is very small and the width of the gap is
comparable to the perfect compounds in figure \ref{fig2}. In the
latter case we see that for the CrAs/CdSe-2 interface the bands
are slightly shifted to higher energy as in a rigid-band model
with respect to the other interfaces but the gap has even larger
width. This is reflected also to the spin-magnetic moments
presented in table \ref{table3} where for this case the Cr
impurity spin moment exceeds the 4 $\mu_B$ while for the other
interfaces it is slightly larger than the perfect systems.

\begin{figure}
\includegraphics[width=\columnwidth]{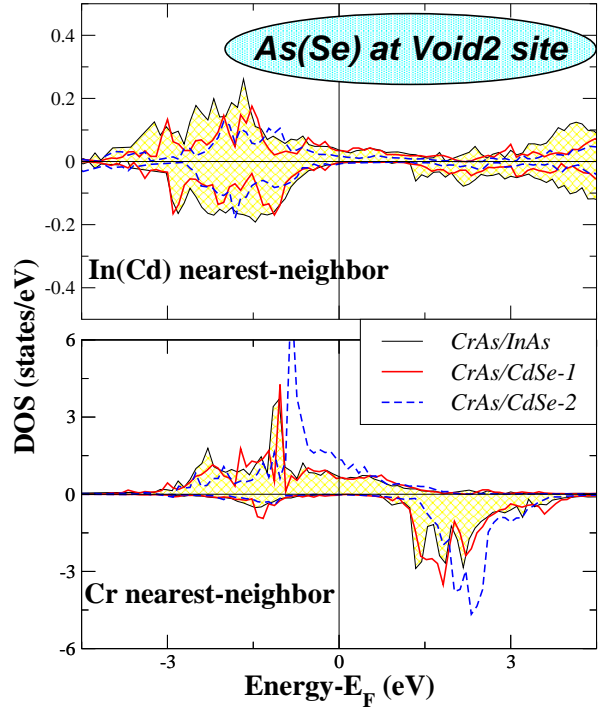}
\caption{(Color online) In the case of As(Se) impurities at Void2
sites we present the atom-resolved DOS of the nearest Cr and
In(Cd) neighbors for all three interfaces. Details as in figure
\ref{fig2}. \label{fig5}}
\end{figure}

\section{As(Se) impuritites}\label{sec4}

Next we will present our results on the As impurities in the case
of the CrAs/InAs and CrAs/CdSe-1 interfaces and on the Se
impurities in the case of the CrAs/CdSe-2 interface. We have
studied all possible cases and in table \ref{table4} we have
concentrated the atom-resolved spin magnetic moments for the
impurities and their nearest neighbors. The DOS of the impurities
in general present no large variation with respect to the As(Se)
atom at the interface or the case of impurities at the perfect
bulk and thus we present only the case of As(Se) atom at Void2
sites, where the As(Se) impurity atom has the same
nearest-neighbors as at the perfect interfacial As(Se) sites. In
all other cases the impurity atom has four As(Se) atoms as nearest
neighbors (two at the interface and two at the CrAs, when the
impurity is located at the Cr or Void1 sites, and two at the
InAs(CdSe) layer, when it is located at the In(Cd) or Void3
sites).

We will first discuss the case of As(Se) impurities at the Cr and
Void1 sites. At the same energy window with the $d$ electrons of
the Cr atoms are located the $p$ electrons of the As(Se) atoms.
The DOS of the impurity atoms is similar to the case of As(Se)
impurities at Cr and Void1 sites in bulk CrAs(CrSe) presented in
\cite{Pouliasis} and thus we do not present them. For the
interfacial As(Se) atoms in the perfect interfaces  the minority
$p$ states of As(Se) are completely occupied leading to small
negative spin moments (see table \ref{table1}). The impurity
As(Se) atoms at Cr or Void1 sites, on the other hand, have four
other As(Se) atoms as nearest neighbors and thus the $p$ states of
the As(Se) impurity atoms have to hybridize with the $p$ states of
the neighboring As(Se) atoms which are almost completely occupied
instead of the Cr $t_2g$ states for which only the majority states
are occupied leading to occupancy of the antibonding $p$ spin-up
states and to positive spin magnetic spin moments as shown in
table \ref{table4}. Due to the reorganization of the charge
induced by the impurity As(Se) states, the number of the occupied
spin-up $p$-states of the neighboring As(Se) atoms increases and,
although the spin magnetic moments remain negative, their absolute
value decreases. In all cases discussed just above, the impurity
states are not located in the gap similar to the what happened for
the same impurities in the bulk systems (see \cite{Pouliasis}) and
the half-metallicity is preserved.

\begin{table}
\centering \caption{Atom-resolved spin magnetic moments in $\mu_B$
for the case of As(Se) impurity atoms at various interfacial sites
in the case of the three interfaces under study. Notation as in
table \ref{table2}. }
 \begin{tabular}{l|c|c|c} \hline \hline
\underline{\bf As(Se) at Cr site}  &  CrAs/InAs & CrAs/CdSe-1 &
CrAs/CdSe-2 \\
As(Se)-imp & 0.108 &   0.118&    0.127 \\
 As(Se) [CrAs]-1st &   -0.243 &   -0.255 &-0.271 \\
  As(Se) [Inter)]-1st & -0.061 &   -0.177 & -0.093 \\ \hline

\underline{\bf As(Se) at Void1 site}  &  CrAs/InAs & CrAs/CdSe-1 &
CrAs/CdSe-2 \\
As(Se)-imp &   0.143 &    0.084 & 0.070 \\
As(Se) [CrAs]-1st &   -0.222 &-0.267 & -0.123 \\
As(Se) [Inter]-1st & -0.114 & -0.208&   -0.087 \\ \hline

\underline{\bf As(Se) at Void2 site}  &  CrAs/InAs & CrAs/CdSe-1 &
CrAs/CdSe-2 \\
As(Se)-imp &    0.269&     0.207&  -0.046\\
 Cr-1st&   3.361&3.212&  3.367\\
  In(Cd)-1st &       0.012&  -0.007&    -0.010\\
As(Se) [CrAs]-2nd&    -0.318&    -0.314&  -0.494\\
 As(Se) [Inter]-2nd&  -0.104&  -0.229&  -0.114\\
  As(Se) [SC]-2nd & 0.056&0.050&  -0.074\\ \hline

\underline{\bf As(Se) at In(Cd) site}  &  CrAs/InAs & CrAs/CdSe-1
& CrAs/CdSe-2
\\ As(Se)-imp & 0.038&  0.269&     0.048\\
 As(Se) [Inter]-1st&-0.128&    -0.013&  -0.584\\
  As(Se) [SC]-1st&  -0.008&0.287&     -0.249\\ \hline

\underline{\bf As(Se) at Void3 site}  &  CrAs/InAs & CrAs/CdSe-1 &
CrAs/CdSe-2\\ As(Se)-imp & 0.093&
0.068&     0.197\\
 As(Se) [Inter]-1st&  -0.069&    -0.142&-0.003\\
  As(Se) [SC]-1st&   -0.009 &   -0.013&    -0.008\\
 \hline \hline
\end{tabular}
\label{table4}
\end{table}

When the As(Se) atoms migrate to Void2 sites, their nearest
neighbors remain two Cr and two In(Cd) atoms as for the perfect
As(Se), but the further neighbors change and the interfacial As
atoms are now next-nearest neighbors and the hybridization between
their $p$ states is intense leading to occupation of the
antibonding spin-up $p$ states and positive values of the spin
magnetic moments of the impurity atoms as shown in table
\ref{table4} for the case of the As impurities in CrAs/InAs and
CrAs/CdSe-1 interfaces or to a large variation of the negative Se
spin moment in the case the CrAs/CdSe-2 interface (the spin moment
of the impurity is half with respect to the Se atom at the perfect
sites). The Cr atoms at the interface have now 5 instead of 4
As(Se) atoms as nearest neighbors but the hybridization of the
$t_{2g}$ $d$-states is little affected and the nearest-neighboring
Cr spin moments are almost identical to the perfect interfaces. A
similar behavior is exhibited by the spin moments of the  As(Se)
second neighbors. The behavior of the spin moments is reflected
also in the DOS presented in figure \ref{fig5}. The DOS of the Cr
nearest neighbors presents for all interfaces under study a very
wide minority-spin gap similar to the perfect compounds with a
large spin-splitting between the occupied majority- and the
unoccupied minority-spin bands. The shape of the Cr DOS is similar
to the Cr atoms at the perfect interfaces shown in figure
\ref{fig2} revealing the small influence of the hybridization with
the impurity As(Se) $p$-states. The same is also valid for the DOS
of the nearest In(Cd) atoms shown also in the same figure although
the gap is smaller for these atoms similarly to what occurred in
the perfect interfaces.

\begin{figure}
\includegraphics[width=\columnwidth]{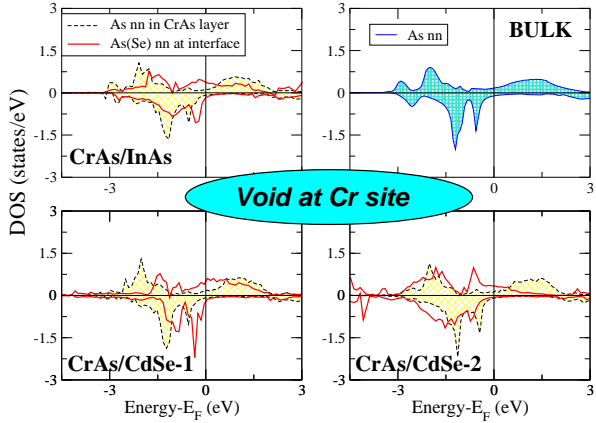}
\caption{((Color online) In the case of Void impurity at an
interfacial Cr site we present the atom-resolved DOS of the
nearest As(Se) neighbors in the CrAs layer and at the interface
for all three interfaces and for the bulk CrAs. Details as in
figure \ref{fig2}. \label{fig6}}
\end{figure}

The case of As(Se) impurities at In(Cd) and Void3 sites is more
difficult to be understood and although the half-metallicity is
preserved the spin moments of the As(Se) impurity atoms and their
nearest As(Se) atoms show large variations. In the case of the
CrAs/InAs interface the As impurity atom has 4 As atoms as nearest
neighbors, in the case of the CrAs/CdSe-1 atom the As impurity
atom has two As at the interface and two Se atoms within the CdSe
layer as nearest neighbors while in the case of the CrAs/CdSe-2
interface the Se impurity atom has four Se atoms as nearest
neighbors. Se atoms have one electron more than the As ones and
thus in general a larger number of spin-up states is occupied as
revealed also from the spin magnetic moments in table
\ref{table1}. Thus especially when the impurity is located at the
In(Cd) site no trend can be found in table \ref{table4} regarding
the spin magnetic moments and the situation is
interface-dependent.

\section{Void impurities}\label{sec5}

\begin{figure}
\includegraphics[width=\columnwidth]{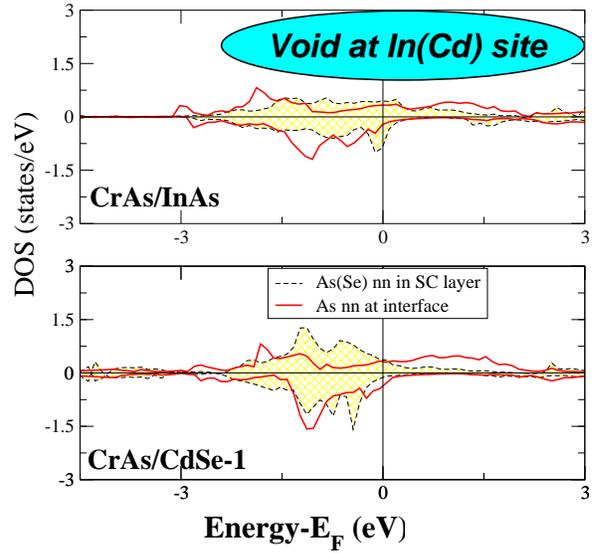}
\caption{(Color online) In the case of Void impurity at an
interfacial In(Cd) site we present the atom-resolved DOS of the
nearest As(Se) neighbors in the semiconducting layer and at the
interface for two of the studied interfaces. Details as in figure
\ref{fig2}. \label{fig7}}
\end{figure}

Up to now we have studied the case of the Cr and As(Se) impurities
and in all cases the half-metallicity was preserved. In this
section we will study the effect of Voids. Voids are Schottky-type
defects with large formation energies. A Void impurity in reality
means that an atom is missing leading to a reorganization of the
charge of the neighboring atoms. We will start our discussion from
the case of a Void at a Cr site and in figure \ref{fig6} we
present the DOS of the nearest As(Se) neighbors for all three
interfaces and for the bulk CrAs. In the case of the bulk the
missing Cr atom means that As atoms have now three instead of four
Cr nearest neighbors and the $p$-bands of As move closer to the
Fermi level without crossing it. In the case of interfaces we
should distinguish between the As atoms within the CrAs layer
which show a behavior identical to the bulk case and the As(Se)
atoms at the interface layer. The latter ones in the perfect
multilayers have two Cr atoms and two In(Cd) as nearest neighbors
and when a Void occurs at a Cr site they lose one of the two Cr
neighbors resulting to a large shift of the $p$-bands which almost
cross the Fermi level. Although one could argue that
half-metallicity is present, a small change of the lattice
constant would lead to loss of the half-metallic character of the
perfect interfaces. As the bands move closer to the Fermi level
the weight of the spin-up states decreases while the bonding
spin-down states are completely occupied leading to larger
absolute values of the negative spin moments of the nearest As(Se)
neighbors with respect to the perfect case as shown in table
\ref{table5}.

\begin{figure}
\includegraphics[width=\columnwidth]{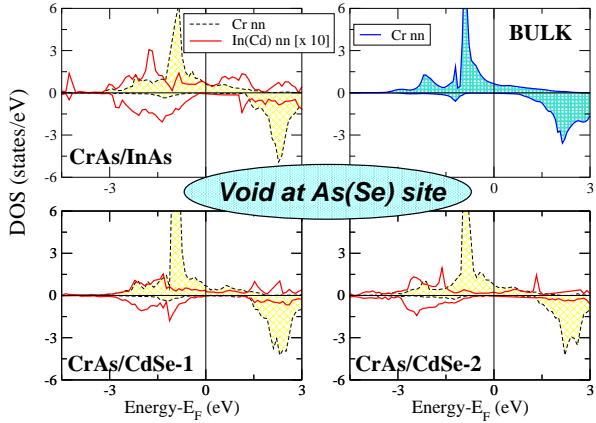}
\caption{(Color online) In the case of Void impurity at an
interfacial As(Se) site we present the atom-resolved DOS of the
nearest Cr and In(Cd) neighbors in the CrAs layer for all three
interfaces and for the bulk CrAs. Note that the In(Cd) DOS has
been multiplied by a factor of 10. Details as in figure
\ref{fig2}.\label{fig8}}
\end{figure}

The same phenomenon occurs also when the Void is located at an
In(Cd) site as shown in figure \ref{fig7}. The missing In(Cd)
neighbors lead to a shift of the $p$-bands of the
nearest-neighboring As(Se) atoms towards higher energies and now
clearly the spin-down band crosses the Fermi level destroying the
half-metallicity. Especially in the case of the CrAs/InAs
interface a spin-down pick is pinned exactly at the Fermi level.
In the case of the CrAs/CdSe-1 interface the bands are more
narrow; the spin-down pick is located below the Fermi level but a
small contraction of the lattice constant could lead to the
pinning of the Fermi level at this pick. As discussed for the case
of Void impurities at Cr sites, also here the absolute values of
the negative spin moments of the neighboring As(Se) atoms increase
(see table \ref{table5}).

\begin{table} \centering \caption{Atom-resolved spin magnetic moments in $\mu_B$
for the case of Void impurities at interfacial Cr, As(Se) and
In(Cd) sites in the case of the three interfaces under study.
Notation as in table \ref{table2}.}
 \begin{tabular}{l|c|c|c} \hline \hline
\underline{\bf Void at Cr site}   & CrAs/InAs &
CrAs/CdSe-1 & CrAs/CdSe-2 \\
 Void-imp &
-0.097& -0.129& -0.061\\ As [CrAs]-1st&  -0.592&
-0.606& -0.613\\ As(Se) [Inter]-1st& -0.350& -0.644& -0.159\\
\hline
 \underline{\bf Void at Ae(Se) site}   & CrAs/InAs &
CrAs/CdSe-1 & CrAs/CdSe-2 \\
Void-imp &  0.201&    0.154&    0.190\\
 Cr-1st & 3.748&    3.691&    3.812\\
  In(Cd)-1st &0.033& -0.008&   0.017 \\ \hline
\underline{\bf Void at In(Cd) site}   & CrAs/InAs &
CrAs/CdSe-1 & CrAs/CdSe-2 \\ Void-imp & -0.094&   -0.073&   0.066\\
 As(Se) [Inter]-1st &
-0.468&   -0.535&   -0.210\\
 As(Se) [SC]-1st   & -0.141&
-0.115& -0.127 \\
 \hline \hline
\end{tabular}
\label{table5}
\end{table}

The last possible case which can occur is when the Void is located
at the interface As(Se) site. The loss of the As(Se) atom only
marginally affects the DOS of the Cr atoms at the interface which
now have three instead of four neighboring atoms. As shown in
figure \ref{fig8} Cr nearest-neighbors at the interface show
similar behavior as in the bulk and the weight of the occupied
minority-spin states is vanishing and the half-metallicity remains
robust. The In(Cd) nearest-neighbors exhibit also a large gap as
shown in the same figure where the DOS of the In(Cd) has been
multiplied by 10 to be visible with respect to the Cr nearest
neighbors. The loss of its As(Se) neighbor leads to an increase of
the local charge of the Cr nearest-neighbors (in a way the regain
the charge which was participating at the bonds with the missing
As(Se) atom). This extra local charge occupies spin-up states
leading to and increase of the Cr spin moment with respect to the
perfect interfaces by about 0.4-0.5 $\mu_B$.

\begin{table} \centering \caption{Atom-resolved spin magnetic moments in $\mu_B$
for the case of In(Cd) impurities at various interfacial sites in
the case of the three interfaces under study. Notation as in table
\ref{table2}.}
 \begin{tabular}{l|c|c|c} \hline \hline
\underline{\bf In(Cd) at Cr site}  & CrAs/InAs &
CrAs/CdSe-1 & CrAs/CdSe-2 \\
In(Cd)-imp &  <0.001&  -0.070&    -0.025\\
 As(Se) [CrAs]-1st&
-0.357&    -0.459&  -0.434\\ As(Se) [Inter]-1st&   -0.146& -0.376&
-0.074\\ Cr [CrAs]-3rd&     3.299&     3.456&  3.173\\
Cr [Inter]-3rd&  3.321&     3.176&     3.464\\ In(Cd) [Inter]-3rd
& -0.014&  -0.048&  -0.006 \\ \hline

\underline{\bf In(Cd) at Void1 site}  & CrAs/InAs & CrAs/CdSe-1 &
CrAs/CdSe-2 \\ In(Cd)-imp  &  0.068 &  0.004 &     0.033\\ As(Se)
[CrAs]-1st &  -0.238 &    -0.318 &  -0.283\\ As(Se) [Inter]-1st &
-0.130 &  -0.269 &    -0.059\\ Cr [CrAs]-2nd &  3.429 &  3.571 &
3.590\\  Cr  [Inter]-2nd &  3.427    & 3.272 & 3.262 \\In(Cd)
[Inter]-2nd & -0.002 & -0.032&    0.004\\ \hline

\underline{\bf In(Cd) at As(Se) site}  & CrAs/InAs & CrAs/CdSe-1 &
CrAs/CdSe-2
\\ In(Cd)-imp & 0.100& 0.148&    0.188\\ Cr-1st& 3.590&
3.590&    3.746\\ In(Cd)-1st&  -0.024&   -0.001& 0.019\\ As(Se)
[CrAs]-3rd -0.414&   -0.421&   -0.416\\ As(Se) [Inter]-3rd
&-0.209& -0.384&   -0.096\\ As(Se) [SC]-3rd& -0.040& -0.032&
-0.023\\ \hline

\underline{\bf In(Cd) at Void2 site}  & CrAs/InAs & CrAs/CdSe-1 &
CrAs/CdSe-2 \\ In(Cd)-imp &  -0.041&  0.088&     0.063\\ Cr-1st &
3.185&  3.256&     3.446\\ In(Cd)-1st&   -0.006&    -0.020&
-0.007\\ As(Se) [CrAs]-2nd&  -0.398&    -0.362&    -0.475\\ As(Se)
[Inter]-2nd &  -0.176&  -0.288&    -0.102\\ As(Se) [SC]-2nd&
-0.044&  -0.039&  -0.067\\ \hline

\underline{\bf In(Cd) at Void3 site}  & CrAs/InAs & CrAs/CdSe-1 &
CrAs/CdSe-2
\\ In(Cd)-imp & 0.058& -0.008&   0.045\\ As(Se) [Inter]-1st&
-0.073&   -0.230& -0.016\\ As(Se) [SC]-1st &-0.011&
-0.028&   -0.011 \\Cr [Inter]-2nd&   3.473& 3.286& 3.638\\
In(Cd) [Inter]-2nd&  0.001&    -0.031& 0.004\\  In(Cd) [SC]-2nd &
-0.003&   0.002&    -0.038\\

 \hline \hline
\end{tabular}
\label{table6}
\end{table}

\section{In(Cd) impurities}\label{sec6}

\begin{figure}
\includegraphics[width=\columnwidth]{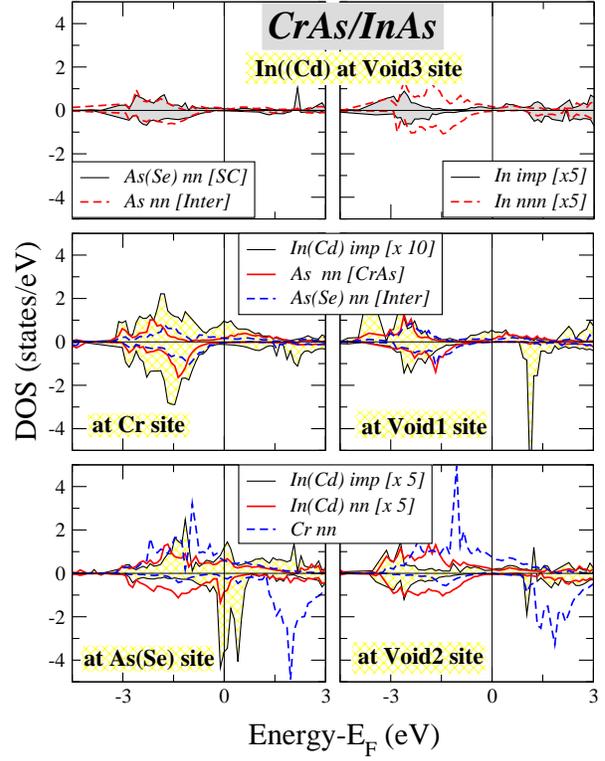}
\caption{(Color online) In the case of In(Cd) impurities at
various interfacial sites, we present the atom-resolved DOS of the
impurity atom and its neighbors. Note that the In(Cd) DOS has been
multiplied by a factor of 5 or 10 in all cases. Details as in
figure \ref{fig2}. \label{fig9}}
\end{figure}

Finally, in the last section we will present our results
concerning the case of In, for the CrAs/InAs interface, and Cd,
for both CrAs/CdSe interfaces, impurities at various sites. All
three interfaces show similar behavior and thus in figure
\ref{fig9} we present the DOS for all possible In impurities for
the CrAs/InAs multilayer. We should note that with respect to the
conservation of the half-metallicity this is the most interesting
case since for the other two CrAs/CdSe interfaces the
half-metallic character is conserved for all cases under study. In
table \ref{table6} we have gathered the atom-resolved spin moments
for all cases under study and as it can be easily deduced from the
table the variation of the spin moments for the same position of
the In(Cd) impurity is similar for all three interfaces and thus
we will restrict our discussion to the CrAs/InAs case.

We expect that the most frequent case to occur would be the In
impurity at the Cr site since such an impurity does not disrupt
the zinc-blende structure. In atoms have only two valence
electrons occupying the deep-energy-lying $s$-states and thus for
the energy window which we examine the $p$-states, which we
observe, have their origin at the nearest As neighbors whose
$p$-states penetrate in the In sites (Cd has only one valence
$s$-electron). Thus the In impurity acts similarly to a Void,
although it does not lead to such large reorganization of the
charge of the neighboring atoms,  leading to slightly larger spin
moment of the neighboring atoms with respect to the perfect
interfaces as shown in table \ref{table6}. Due to the small weight
of the In $p$-states we have multiplied the corresponding DOS with
a factor 5 or 10 in figure \ref{fig9} to make it visible. With
respect to the case of Void impurity at the Cr site, here the
shift of the bands of the nearest-neighboring As atoms is smaller
keeping the half-metallic character of the interface although the
gap is considerably shrinking.

When the In impurity is located at the Void1 site, the disturbance
of the lattice is smaller with respect to the case just presented,
although both Cr and Void1 sites have the same nearest-neighbors
and as shown in figure \ref{fig9} the width of the gap remains
unchanged. Due to the negligible weight of the In $p$ states also
the occurrence of In impurities at Void2 and Void3 sites leads to
a slight variation of the spin moments but the half-metallicity is
preserved and the gap retains a large width. To conclude we should
discuss also the case of In impurities at As sites. As atoms at
the interface have two Cr atoms as nearest neighbors and the
hybridization between the As $p$- and Cr $t_{2g}$-orbitals is
strong. The substitution of an As atom by an In one leads to
reduced hybridization for the Cr orbitals and Cr atoms at the
interface regain the charge participating at the bonds with the
missing As atom. This extra charge is accommodated at the Cr
spin-up states leading to larger spin magnetic moments of the Cr
atoms at the interface which now are about 3.59 $\mu_B$ instead of
3.25 $\mu_B$ in the case of the perfect CrAs/InAs interface
presented in table \ref{table1}. The In impurity atom and its
nearest-neighboring In atoms have states within the gap, as shown
in figure \ref{fig9} but if we take into account that we have
multiplied the In DOS by 10 their real weight at the Fermi level
is negligible with respect to the Cr majority-spin DOS. Thus we
can safely consider that the half-metallicity is conserved
although as shown by the Cr DOS, the gap in the minority-spin band
seriously shrinks and the Fermi level is near the right edge of
the gap.

\section{Conclusion}\label{sec7}

We have studied using the Korringa-Kohn-Rostoker method the
appearance of single impurities at interfaces between the
half-metallic ferromagnet CrAs and the binary InAs and CdSe
semiconductors. In the case of bulk CrAs studied in reference
\cite{Pouliasis} we had shown that most impurities affect the
half-metallic character of CrAs inducing states either at the
edges of the gap or in the middle of the gap. But multilayers are
very thin as experiments show \cite{Mizuguchi}
 and thus we cannot use the impurity calculations for the bulk to
 discuss the case of interfaces. We have studied al possible
 single impurities at the interfaces. Contrary to the bulk systems
 almost all defects were not affecting the half-metallic character
 of the perfect interfaces. The only exception were Void
 impurities at Cr or In(Cd) sites. The missing Cr or In(Cd) atom
 leads to a reorganization of the charge of the surrounding atoms
 and as a result the $p$  bands of the nearest neighboring As(Se)
 atom shift to higher energies crossing the Fermi level and
 leading to loss of the half-metallicity. This is the opposite
 behavior that the one exhibited by the Void impurities at Cr site in the
 bulk CrAs alloy where the DOS remained almost unchanged
 \cite{Pouliasis}. The different behavior of the Void impurities
 should be attributed to the lower dimensionality of the
 interfaces with respect to the bulk. But Void impurities are
 Schottky-type and we expect them to exhibit high formation
 energies and thus their number should be small in the
  experimental systems. Thus contrary to the bulk CrAs and
  eventually thick films showing a bulk-like behavior,
 impurities are expected to affect only marginally the
 half-metallic character  of the interfaces in the case of thin multilayers
and the latter ones are promising for spintronic devices.


\end{document}